\documentclass[a4paper,11pt]{article}
\usepackage[english]{babel}
\usepackage[dvips]{graphicx}
\usepackage{amssymb,amsfonts,amsmath,graphicx,subfigure}
\usepackage{graphics,graphicx,epsfig}

\begin{document}

\title{A Multifractal Detrended Fluctuation Description of Iranian Rial-US Dollar Exchange Rate  }
\author{P. Norouzzadeh  \footnote{e-mail: noruzzadeh@farda.ir}   \\
Quantitative Analysis Research Group,\\ Farda Development
Foundation, Tehran, Iran \\} \maketitle

\begin{abstract}
The miltifractal properties and scaling behaviour of the exchange
rate variations of the Iranian rial against the US dollar from a
daily perspective is numerically investigated. For this purpose the
multifractal detrended fluctuation analysis (MF-DFA) is used.
Through multifractal analysis, the scaling exponents, generalized
Hurst exponents, generalized fractal dimensions and singularity
spectrum are derived. Moreover, contribution of two major sources of
multifractality, that is, fat-tailed probability distributions and
nonlinear temporal correlations are studied.
\end{abstract}

%PACS numbers: 05.40; 89.90 +n \\

\section{Introduction}

For more than two decades, there has been considerable interest in
the investigation of the scaling behaviour on fractal models. The
pioneering work \cite{Mandelbrot} on fractals introduced the concept
of fractals and showed some relation between self-similar fractals
and self-affine fractals. Self-affine fractals \cite{Vicsek,
Barabasi, Halsey, Paladin} constitute random and complicated
structure and have been applied to a broader range of problems, such
as the Eden and ballistic deposition model \cite{Family, Jullien,
Freche, Meakin} , mountain heights, clouds, coast lines, and cracks.
Among other examples of many fractal models, the self-avoiding
random walk, random resistor, polymer bonds, turbulences, chaotic
motions can be mentioned \cite{Vicsek, Halsey, Paladin, Lee, Tel,
Farmer, Benzi} , \emph{etc}. Specially, the real data from different
financial markets show apparent multifractal properties
\cite{Pasquini, Ivanova, Bershadskii, Di Matteo, Fisher, Vandewalle,
Bershadskii2, Matia, Drozdz}. Moreover,recently, financial analysis
of foreign exchanges has became one of the outstanding topics in
econophysics \cite{Takayasu}. Many of these researches apply
multifractal analysis framework as a basic framework. In universal
multifractal framework, the statistics of the data are fully
described with some parameters, taking into account two
complementary aspects of financial time series: the multiple scaling
and the Pareto probability distributions, which is a generic feature
of
multifractal processes \cite{Schmitt}.\\
It has been shown that there are two main factors leading to
multifractal behaviour of financial time series, nonlinear time
correlations between present and past events and the heavy-tailed
probability distributions of functions. Based on ref. \cite{Matia} ,
for the stocks, the main contribution to multifractality comes from
a broad distribution of returns while a long memory present in this
kind of data contributes only marginally. It should be noted that
the nature of correlations leading to the multifractal dynamics of
the variations is strongly nonlinear and, curiously, cannot be
simply related to some well-known correlation type like a slowly
decreasing volatility autocorrelation with an imposed daily pattern.
For example, one even has to consider the nonlinear dependencies in
the volatility itself in order to reveal how the temporal
correlations contribute to multifractality in the stock
market and foreign exchange data.\\
In this paper, the rial-dollar exchange rate data is studied with
the focus on their fractal properties. The multifractal detrended
fluctuation analysis is applied which is a well-established method
of detecting scaling behaviour of time series. In Section 1,
theoretical backgrounds including MFDFA method, sources of
multifractality, multifractality finger prints and strength of
multifractality are reviewed. Data are described in Section 2.
Numerical results are presented in Section 3 and finally,
conclusions are given in Section 4.

\section{Theoretical backgrounds}
\subsection{Method}
Detrended fluctuation analysis (DFA) is a scaling analysis
technique providing a simple quantitative parameter-the scaling
exponent $\alpha$-to represent the correlation properties of a
time series \cite{C.K. Peng}. The advantage of DFA over many
techniques are that it permits the detection of long-range
correlations embedded in seemingly non-stationary time series, and
also avoids the spurious detection of apparent long-range
correlations that are an artifact of non-stationarity.
Additionally, the advantages of DFA in computation of $H$ over
other techniques (for example, the Fourier
transform) are:\\
\begin{itemize}
\item inherent trends are avoided at all time scales; \item local
correlations can be easily probed.
\end{itemize}

To implement the DFA, let us suppose there is a time series, $N(i)
(i=1,...,N_{max})$. The time series $N(i)$ is integrated:
\begin{equation}
y(j)=\sum_{i=1}^{j}[N(i)-\langle N \rangle]
\end{equation}
where:
\begin{equation}
\langle N\rangle=\frac{1}{N_{max}}\sum_{j=1}^{N_{max}}N(i).
\end{equation}
Next $N(i)$ is broken up into $K$ non-overlapping time intervals,
$I_{n}$, of equal size $\tau$ where $n=0,1,...K-1$ and $K$
corresponds to the integer part of $N_{max}/\tau$. In each box,
the integrated time series is fitted  by using a polynomial
function, $y_{pol}(i)$, which is called the local trend. For
order-\emph{l} DFA (DFA-1 if $l$=1, DFA-2 if $l$=2, etc.), the
\emph{l}-order polynomial function should be applied for the
fitting. The integrated time series $y(i)$ is detrended in each
box, and calculated the detrended fluctuation function:
\begin{equation}
Y(i)=y(i)-y_{pol}(i).
\end{equation}
For a given box size $s$, the root mean square fluctuation is
calculated:
\begin{equation}
F(s)=\sqrt{\frac{1}{N_{max}}\sum_{i=1}^{N_{max}}[Y(i)]^{2}}
\end{equation}
The above computation is repeated for box sizes $s$ (different
scales) to provide a relationship between $F(s)$ and $s$. A power
law relation between $F(s)$ and $s$ indicates the presence of
scaling: $F(s)\sim s^{\alpha}$. The parameter $\alpha$, called the
scaling exponent or correlation exponent, represents the
correlation properties of the signal: if $\alpha=0.5$, there is no
correlation and the signal is an uncorrelated signal \cite{C.K.
Peng}; if $\alpha<0.5$, the signal is anticorrelated; if
$\alpha>0.5$, there are positive correlations in the signal. In
the two latest cases, the signal can be well approximated by the
fractional Brownian motion law \cite{J. Feder}.\\
For a further characterization of data it is meaningful to extend
Eq. (15) by considering the more general fluctuation functions
\cite{Barabasi2}. Simply, it is achieved by averaging over all
boxes to obtain the $q$th order fluctuation function
\begin{equation}
F_{q}(s)=[\frac{1}{2N_{max}}\sum_{i=1}^{N_{max}}(F^{2}(s))^{q/2}]^{1/q},
\end{equation}

where, in general, the index variable $q$ can take any real values
except zero. If the analyzed signal develops fractal properties,
the fluctuation function reveals power-law scaling
\begin{equation}
F_{q}(s)\sim s^{h(q)}
\end{equation}
for large $s$. The scaling exponents $h(q)$ can be then obtained by
observing the slope of log-log plots of $F_{q}$ vs.$s$. The family
of the exponents $h(q)$ describe the scaling of the $q$th order
fluctuation function. For positive values of $q$, $h(q)$ exponents
describe the scaling behaviour of boxes with large fluctuations
while those of negative values of $q$, describe scaling behaviour of
boxes with small fluctuations \cite{Havlin}. For stationary time
series, the exponent $h(2)$ is identical to the Hurst exponent. Thus
the exponents $h(q)$ are called as the generalized Hurst exponents
\cite{Havlin}. For monofractal time series which are characterized
by a single exponent over all scales, $h(q)$ is independent of $q$,
whereas for a multifractal time series, $h(q)$ varies with $q$. This
dependence is considered to be a characteristic property of
multifractal processes \cite{Havlin}. The $h(q)$ obtained from
MF-DFA is related to the Renyi exponent $\tau(q)$ by
\begin{equation}
qh(q)=\tau(q)+1.
\end{equation}
Therefore, another way to characterize a multifractal series is the
singularity spectrum $f(\alpha)$ defined by \cite{J. Feder}
\begin{equation}
\alpha=h(q)+qh'(q), \hspace{1cm} f(\alpha)=q[\alpha-h(q)]+1,
\end{equation}
where $h'(q)$ stands for the derivative of $h(q)$ with respect to
$q$. $\alpha$ is the H$\ddot{o}$lder exponent or singularity
strength which characterizes the singularities in a time series.
The singularity spectrum $f(\alpha)$ describes the singularity
content of the time series. Finally, it must be noted that $h(q)$
is different from the generalized multifractal dimensions
\begin{equation}
D(q)\equiv \frac{\tau(q)}{q-1}=\frac{qh(q)-1}{q-1},
\end{equation}
that are used instead of $\tau(q)$ in some papers. While $h(q)$ is
independent of $q$ for a monofractal time series with compact
support, $D(q)$ depends on $q$ in that case.

\subsection{Sources of multifractality}
Generally, there are two different types of sources for
multifractality in time series: (i) due to different long-range
temporal correlations for small and large fluctuations, and (ii) due
to fat-tailed probability distributions of variations. Both of them
need a multitude of scaling exponents for small and large
fluctuations. Two procedure is followed to find the contributions of
two sources of multifractality and to indicate the multifractality
strength: (i) shuffling, and (ii) phase randomization. Shuffling
procedure preserves the distribution of the variations but destroys
any temporal correlations. In fact, one can destroy the temporal
correlations by randomly shuffling the corresponding time series of
variations. What then remains are data with exactly the same
fluctuation distributions but without memory.The shuffling procedure
consists of the following steps

\begin{itemize}

\item[{(i)}] Generate pairs $(p,q)$ of random integer numbers
(with $p,q \leq N$) where $N$ is the total length of the time
series to be shuffled.

\item[{(ii)}] Swap entries $p$ and $q$.

\item[{(iii)}] Repeat two above steps for $20 N$ times. (This step
ensures that ordering of entries in the time series is fully
shuffled.)

\end{itemize}
In order to study the contribution of the fat-tailed variations on
the multifractality, the surrogate data are used. In fact, the
non-Gaussianity of the distributions can be weakened by creating the
phase-randomized surrogates \cite{Theiler}. The Phase randomization
steps are:
\begin{itemize}
\item[{(i)}] Take discrete Fourier transform of time series.\\
\item[{(ii)}] Multiply the discrete Fourier transform of the data by random phases.\\
\item[{(iii)}] Perform an inverse Fourier transform to create a
phase randomized surrogates.
\end{itemize}
Phase randomization preserves the amplitudes of the Fourier
transform but randomizing the Fourier phases. This procedure
eliminates nonlinearities, preserving only the linear properties
of the original time series \cite{Panter}.
\subsection{Multifractality finger prints}
One can see that in the whole q-range the generalized Hurst
exponents $h(q)$ can be fitted well by the formula
\begin{equation}
h(q)=\frac{1}{q}-\frac{ln[a^{q}+b^{q}]}{qln2}
\end{equation}
which corresponds to $\tau(q)=-ln[a^{q}+b^{q}]/ln2$ . This formula
can be obtained from a generalized binomial multifractal model
\cite{Koscielny}. Instead of choosing $a$ and $b$, the Hurst
exponent $h(1)$ and the persistence exponent $h(2)$ could be
chosen. From knowledge of two moments, all the other momemts
follow. Here the formula is used only to show that the infinite
number of exponents $h(q)$ can be described by only two
independent parameters, $a$ and $b$. These two parameters can then
be regarded as multifractal finger prints for a considered time
series.
\subsection{Strength of multifractality}
In the generalized binomial multifractal model, the strength of
the multifractality of a time series can be characterized by the
difference between the maximum and minimum values of $\alpha$,
$\alpha_{max}-\alpha_{min}$. When $q\frac{dh(q)}{dq}$ approaches
zero for $q$ approaching $\pm\infty$, then
$\triangle\alpha=\alpha_{max}-\alpha_{min}$ is simply given by
\begin{equation}
\bigtriangleup\alpha=h(-\infty)-h(\infty)=\frac{ln(b)-ln(a)}{ln2}.
\end{equation}
It must be noted that this parameter is identical to the width of
the singularity spectrum $f(\alpha)$ at $f=0$. The wider
singularity spectrum the richer multifractality.

\section{Data Analysis}
The data which is analyzed are the time series of the daily closing
exchange rate logarithmic variations (that is, ln(P(t))/ ln(P(t+1))
for the time period 24th September 1989, to 15th November 2003. So
that our database consists of 4369 exchange rates and 4368 daily
variations. The sources of this data is the central bank of the
islamic republic of Iran. In Fig. 1 a time series corresponding to
daily values of the Iranian rial-US dollar exchange rates in
mentioned period is presented. A great increment in dollar price is
seen about 16th May 1995 because of Iranian government decision on
lifting the ban on foreign exchanges price variations. The Iranian
government has managed to keep the exchange rate stable at around
8000 rials per US dollar ever since March 2000. Also, Table 1
provides summary statistics of logarithmic variations of exchange
rates.
\begin{table}[htb]
\begin{center}
\caption{\label{Tb2}Mean, standard deviation, skewness, and kurtosis
of rial-dollar exchange rate variations.}

\medskip
\begin{tabular}{cccc}
\hline\hline $Mean$&$Std. Dev.$&$Skewness$&$Kurtosis$\\\hline
0.00055 & 0.0117 & -1.2504 & 49.925 \\
\hline\hline
\end{tabular}
\end{center}
\end{table}
According to data in Table 1, a negatively large skew is seen. The
probability distribution function of variations also show a high
degree of peakedness and fat tails relative to a normal
distribution. Thus there is a clear departure from Gaussian
normality. The departure from a Gaussian Cumulative Distribution
Function (CDF) can be clearly seen in Fig. 2, where the CDF of
variations against a Gaussian CDF is depicted.

\section{Results}
The fluctuation functions $F_{q}(s)$ for timescales ranging from 3
days to $N$/5 are calculated, where $N$ is the total length of the
time series, and for $q$ varying between -10 and 10, with a step of
0.5. Fig. 3 shows the MF-DFA2 fluctuations $F_{q}(s)$ for various
$q$'s.\\
A crossover with great magnitude (like as a phase transition) in
fluctuation function is seen for negative $q$ values in the range
$30<s<65$. The position of crossover doesn't have sensitivity to
decreasing or increasing $q$ values. The only interest behaviour is
the asymptotic behaviour of $F_{q}(s)$ at large times $s$. One can
clearly observe that above the crossover region, the $F_{q}(s)$
functions are straight lines in the double logarithmic plot, and the
slopes increase slightly when going from high positive moments
towards high negative moments (from the top to the bottom in Fig. 3).\\
For the sake of better studying the large fluctuations, randomized
data ( both of reshuffled and surrogate data) have been used. Fig. 4
indicates that, the magnitude of change in crossover for reshuffled
data is very large relative to the surrogate data. In fact, one can
say that such an effect originates mainly from temporal
correlations. Moreover, the position of crossover is intended to
left (about $s\simeq 4$) because of randomizing . \\
Monofractal time series are associated with a linear plot $\tau(q)$,
while multifractal ones possess the spectra nonlinear in $q$. The
highest nonlinearity of the spectrums, the strongest multifractality
in time series. Calculations indicate that the time series of
exchange rate variations can be of multifractal nature. In order to
visualize the scaling character of the data, in Fig. 5, the
corresponding multifractal spectra is shown. Fig. 5 shows three
examples of $\tau(q)$ for the original (solid), surrogate (dotted)
and reshuffled (dashed) data. The nonlinearity of $\tau(q)^{,}s$ is
much weaker for the modified time series than for the original ones.
Additionally, surrogate data show less nonlinearity based on Fig. 5
and therefore, their contribution to multiscaling relative to
reshuffled data is less.\\
The $h(q)$ spectra has been fitted in the range $-10\leq q \leq10$
for original, reshuffled and surrogate series by Eq. (10).
Representative example for original series is shown in Fig. 6. The
dotted line in Fig. 6 is obtained by best fits of $h(q)$ by Eq.
(10). The respective parameters $a$ and $b$ for original, reshuffled
and surrogate series are listed in Table 2. It is notable that in
each single case, the $q$ dependence of $h(q)$ for positive and
negtive values of $q$ can be characterized very well by the two
parameters, and all fits remain within the error bars of the $h(q)$
values.

\begin{table}[htb]
\begin{center}
\caption{\label{Tb2}Multifractality finger prints (parameters a
and b) and strength for original, reshuffled and surrogate data.}

\medskip
\begin{tabular}{|c|c|c|c|}
\hline Time series&a&b&$\bigtriangleup\alpha$\\\hline
Original data & 0.03 & 1.07 & 3.54 \\
Reshuffled data & 0.51 & 0.93 & 0.60\\
Surrogate data& 0.69 & 0.8  & 0.15\\
\hline
\end{tabular}
\end{center}
\end{table}
It is seen that the strength of multifractality in rial-dollar
exchange rate variations is very powerful. Moreover, multifractalty
strength in randomized data decreases specially in
surrogate data based on values in Table 2.\\
In order to visualizing and better understanding the strength of
multifractality for original, reshuffled and surrogate data, the
singularity spectrum of series are shown in Fig. 7. Both the widths
of the $f(\alpha)$ spectra in each randomized data are much smaller
than for the original one. This behaviour of the reshuffled time
series confirms that the persistent autocorrelations play an
important role in multiscaling of the price variations. But, The
spectra for the surrogates are typically much narrower than for the
reshuffled data which can be interpreted as an evidence of the
influence of extremely large non-Gaussian events on the fractal
properties of the time series.
\section{Conclusions}
The multifractal properties of the Iranian rial-US dollar exchange
rate logarithmic variations has been studied in this paper through
multifractal detrended fluctuation analysis. It is shown that the
time series for exchange rate variations exhibit the characteristics
that can be interpreted in terms of multifractality. Its degree
expressed by e.g. the widths of the singularity spectra $f(\alpha)$
indicate a strong multifractality. Moreover, although the most
multifractality of the exchange rate variations data is due to
different long-range correlations for small and large fluctuations,
the shape of the probability distribution function also contributes
to the multifractal behaviour of the time series.

\newpage
\begin{figure}[h]
{\centering
\resizebox*{0.7\textwidth}{0.35\textheight}{\includegraphics{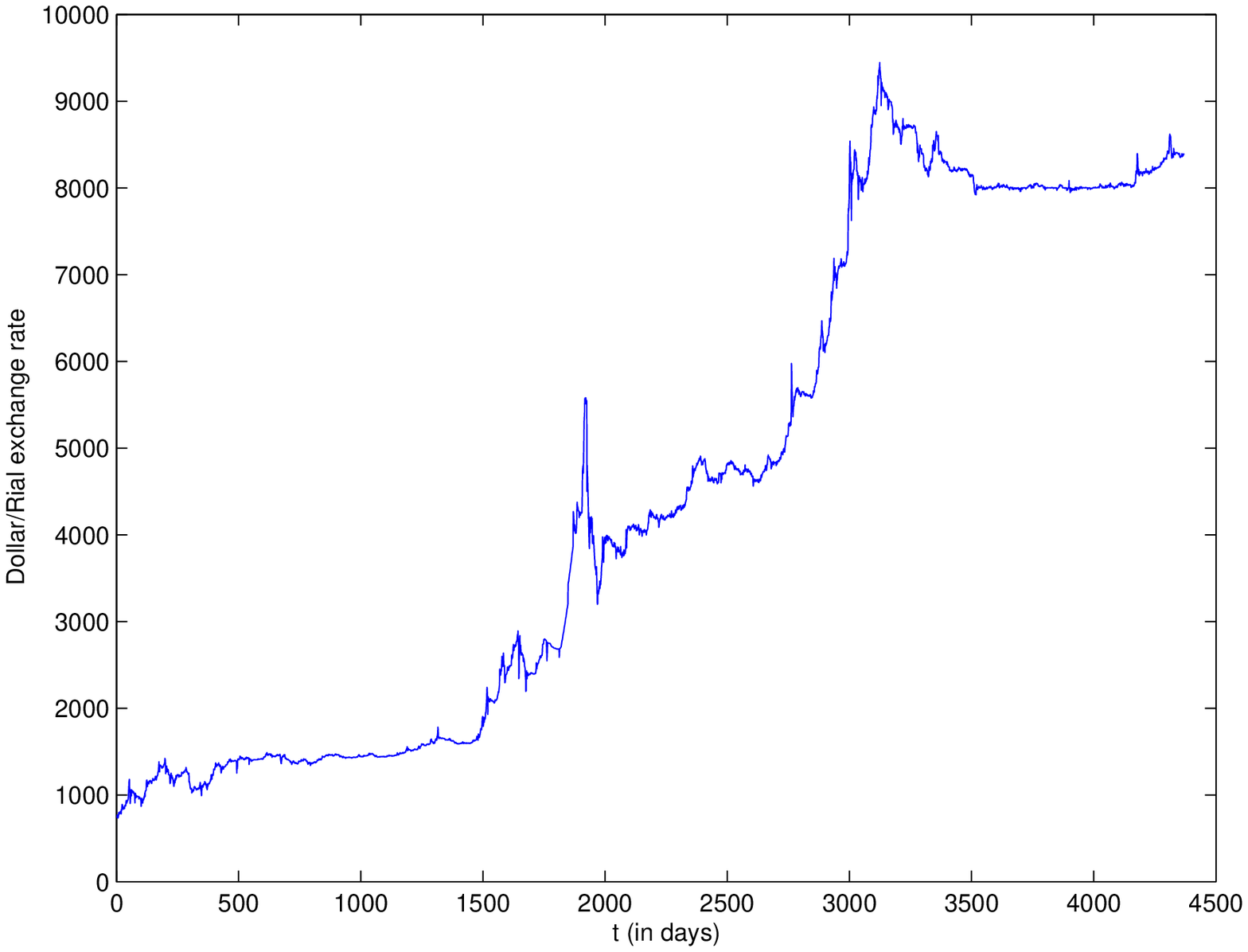}}
\par} \caption{Daily closure rial-dollar exchange rates history (1989-2003).}
\end{figure}
\begin{figure}[h]
{\centering
\resizebox*{0.7\textwidth}{0.35\textheight}{\includegraphics{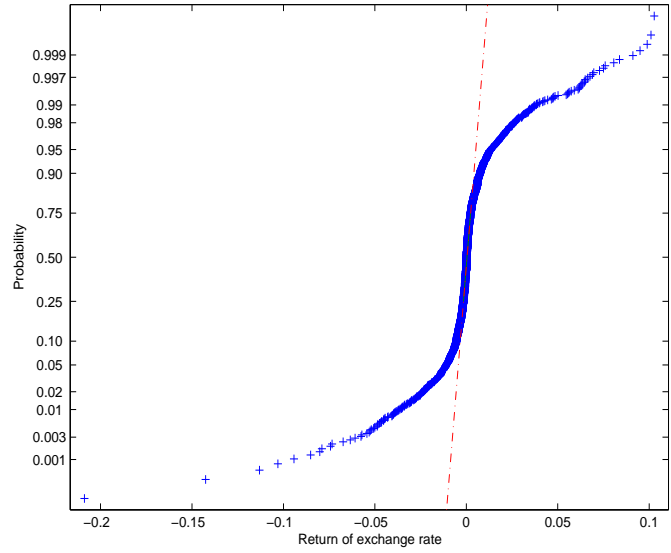}}
\par} \caption{Cumulative distribution function of rial-dollar exchange rate variations against
a Gaussian cumulative distribution}
\end{figure}
\begin{figure}[h]
\begin{center}
\includegraphics[width=12cm]{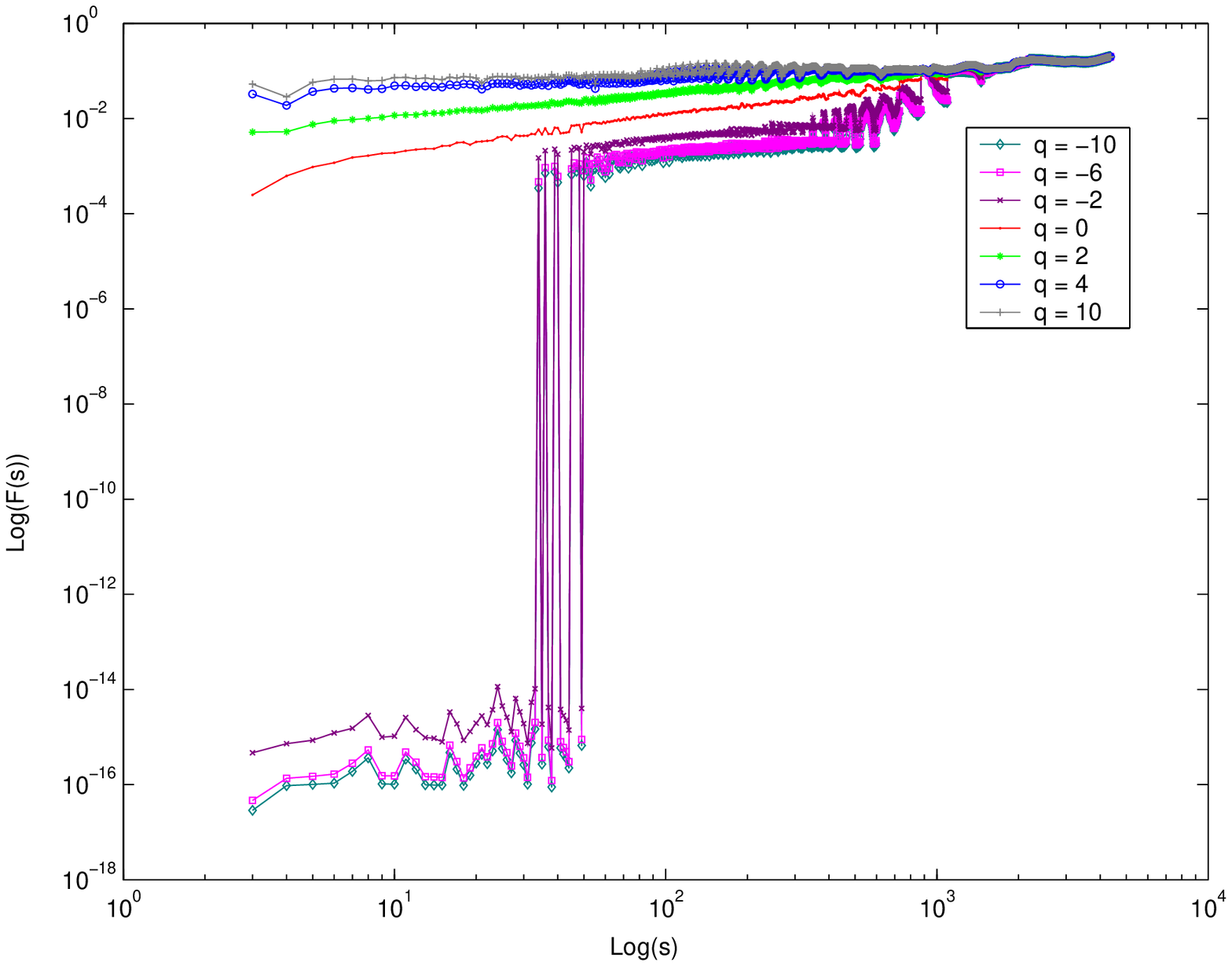}
\end{center}
\caption{The multifractal fluctuation function $F_{q}(s)$ obtained
from multifractal DFA2 for variations of rial-dollar exchange rates
in the period 1989 to 2003.}
\end{figure}
\begin{figure}[h]
\begin{center}
\includegraphics[width=12cm]{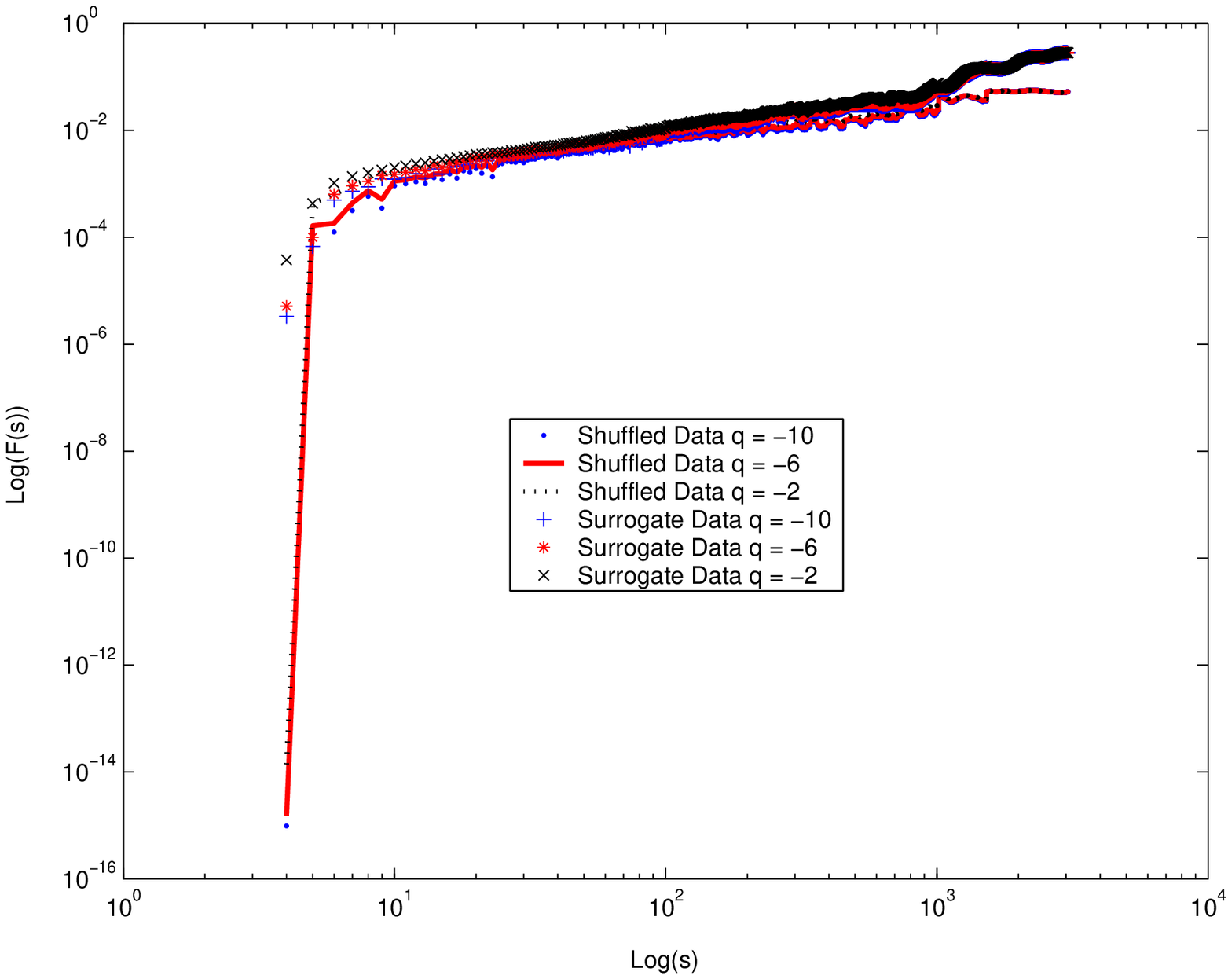}
\end{center}
\caption{The multifractal fluctuation function $F_{q}(s)$ obtained
from multifractal DFA2 for randomized (reshuffled and surrogate)
variations of rial-dollar exchange rates in the period 1989 to
2003.}
\end{figure}
\begin{figure}[h]
{\centering
\resizebox*{0.7\textwidth}{0.35\textheight}{\includegraphics{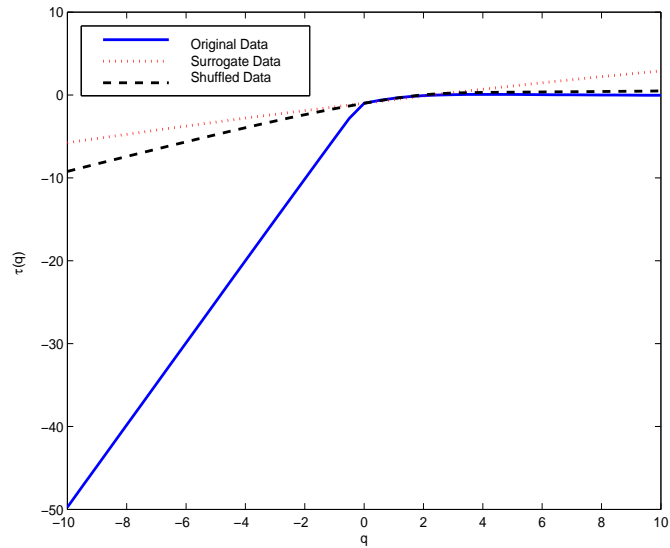}}
\par} \caption{Comparison of the miltifractal spectra $\tau(q)$ of the
original and randomized exchange rate variations: original (solid),
surrogate (dotted) and reshuffled (dashed) time series.}
\end{figure}
\begin{figure}[h]
{\centering
\resizebox*{0.7\textwidth}{0.35\textheight}{\includegraphics{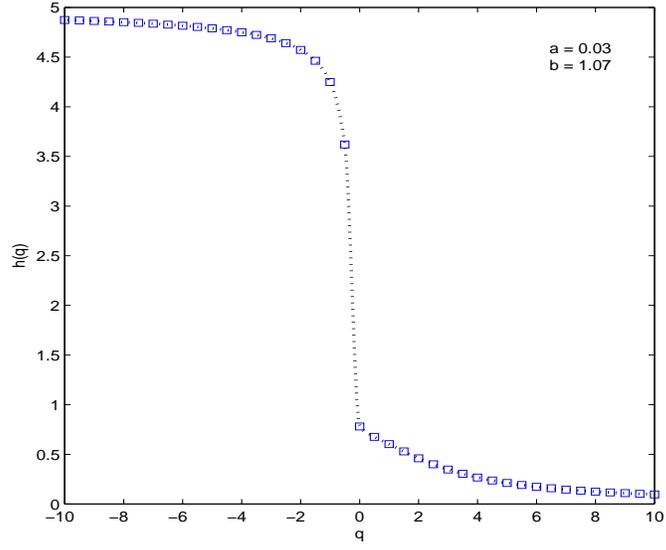}}
\par} \caption{The generalized Hurst exponents $h(q)$ for the rial-dollar
exchange rate variations in period 1989 to 2003. The fitted curve
has been shown by dotted line.}
\end{figure}
\begin{figure}[h]
{\centering
\resizebox*{0.7\textwidth}{0.35\textheight}{\includegraphics{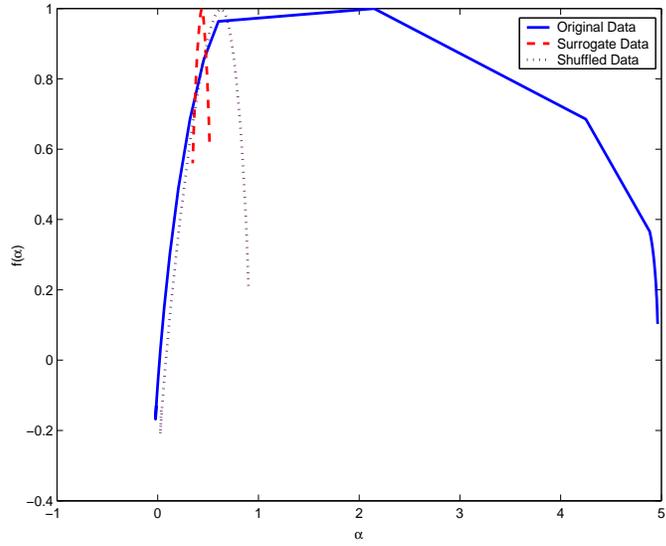}}
\par} \caption{Comparison of the singularity spectra for original and
randomized data: original (solid), reshuffled (dotted) and surrogate
(dashed) time series.}
\end{figure}
\end{document}